\documentclass{sig-alternate-10pt}
\usepackage{epsfig}
\usepackage{graphicx}
\usepackage{balance}
\usepackage{comment}
\usepackage{url}
\usepackage{cite}
\usepackage{textcomp}
\usepackage{listings}
\usepackage{color,soul}
\usepackage{colortbl}
\usepackage[dvipsnames]{xcolor}
\usepackage{leading}

\newcommand{\subparagraph}{} 
\usepackage[noindentafter]{titlesec}

  {\begin{list}{$\bullet$}%
     {\setlength{\parsep}{0pt}%
      \setlength{\topsep}{0pt}%
      \setlength{\itemsep}{2pt}}}%
  {\end{list}}

\newcommand{\customvspace}{\vspace{0in}}

\newcommand\note[1]{} 

\newcommand\candidate[1]{#1} 

\newcommand\candidatetwo[1]{} 

\begin{document}

\title{\LARGE \bf Making I/O Virtualization Easy with Device Files}


\author{
{\large
Technical Report 2013-04-13, Rice University
}
\\
\\
{\large 
     Ardalan Amiri Sani$^{\star}$,
     Sreekumar Nair$^{\dagger}$,
     Lin Zhong$^{\star}$,
     Quinn Jacobson$^{\ddagger}$
}
\\
{\normalsize
     $^{\star}$Rice University, 
     $^{\dagger}$Dynavisor, Inc.,
     $^{\ddagger}$Vibrado Technologies
}
}

\date{} 
\maketitle

\thispagestyle{empty}

\section*{Abstract}
\label{sec:abstract}

Personal computers have diverse
and fast-evolving I/O devices, making their I/O virtualization
different from that of servers and data centers. In this paper, we
present our recent endeavors in simplifying I/O virtualization for
personal computers. Our key insight is that many operating systems,
including Unix-like ones, abstract I/O devices as device files. There is
a small and stable set of operations on device files, therefore,
I/O virtualization at the device file boundary requires a one-time
effort to support various I/O devices.

We present {\em devirtualization}, our design of I/O
virtualization at the device file boundary and its implementation for 
Linux/x86 systems. We are able to virtualize various GPUs, input
devices, cameras, and audio devices with fewer than 4900 LoC,
of which only about 300 are specific to I/O device classes. 
Our measurements show that devirtualized devices achieve interactive
performance indistinguishable from native ones by human users, even when
running 3D HD games.

\section{Introduction}
\label{sec:introduction}

The value of virtualization is increasingly recognized for
personal computers\footnote{By personal computer, we refer to desktops
and mobile computers of diverse form factors including laptops,
smartphones, and tablets.}~\cite{heiser2008role, armand2008shared,
barr2010vmware, Andrus2011, gudeth2011delivering, vmware_mvp, okl4}. As
personal computers are used for diverse purposes, virtualization allows
a user to have multiple virtual machines (or guests) inside the same
computer, each for a dedicated purpose: one for work, one for personal
use, and one for sharing with others~\cite{liu09mobisys}. Also, as
hardware and software of personal computers evolve rapidly,
virtualization allows the legacy code to be reused in new systems.

We are particularly interested in whole system virtualization, which
allows multiple guest operating systems to reside in the same
computer and provides strong isolation between them. In this paper, we
assume a hosted hypervisor, i.e., a hypervisor running inside a host OS,
and assume the guests and the host use the same OS or different versions of
the same OS. 
As a result, we target our solutions for scenarios like having
multiple virtual machines in the same personal computer or reusing
legacy code.

While good solutions exist for CPU and memory
virtualization~\cite{dall2010ols,Kivity2007}, virtualizing I/O devices of
personal computers has proven to be much harder due to their diversity
in function and implementation. To support our targeted scenarios
(above), the I/O virtualization solution
must (\textit{i}) require {\em low development
effort} to support various I/O devices; (\textit{ii}) allow for {\em
sharing the I/O device} between the host and the guests; (\textit{iii})
support {\em legacy devices} that are not specialized for
virtualization; and (\textit{iv}) be {\em portable} to support
virtualization across different
versions of the same OS. The solution should also provide {\em adequate
performance} for personal computers. Unfortunately, available solutions
do not provide one or more of these properties. 

In this paper, we study a novel boundary, \textit{device files}, for I/O
virtualization that meets all the aforementioned properties 
for personal computers. Modern OSes, such as Unix-like
ones employ device files to abstract I/O devices~\cite{UnixFile}. To
virtualize an I/O device, our solution creates a virtual device file in
the guest OS for the corresponding device file in the host. Threads of guest
processes issue file operations to this virtual device file as if it
were the real device file. A thin indirection layer, called
\textit{Common Virtual Driver} (\textit{CVD}), forwards such file
operations to the host to be executed by the \textit{unmodified} host
device driver.


Our use of device files as the boundary for I/O virtualization is
motivated by four properties: 
(\textit{i}) {\em Low development effort:} device files are common to
many important classes of I/O devices in personal computers, including
GPUs, input devices, camera, and audio devices. Moreover, the device
file boundary is narrow due to the small set of file operations.
For example, Linux has about 30 file operations, and only about 10
of them are used by most I/O devices. Finally, since device files are at
a higher layer than device drivers, virtualization at this boundary
allows for reuse of the device drivers in the host.
(\textit{ii}) {\em Sharing:} virtualization at the device file boundary
readily supports sharing the device between the host and the
guests. If multiple host applications can use the same device file in the
host OS, then guest applications can use the same device file as well.
(\textit{iii}) {\em Legacy support:} device files are used by existing
devices in personal computers; therefore, virtualization at this
boundary can support these devices.
(\textit{iv}) {\em Portability}: the device file boundary has been quite
stable across different versions of mature OSes such as Linux.

We present our design of the CVD and its implementation
for Linux/x86 computers, called \textit{devirtualization}, which
realizes the theoretical benefits analyzed above and achieves {\em
performance adequate for personal computing}. We have
addressed a fundamental challenge that \textit{the guest and the host
reside in different virtualization domains, creating a barrier for
forwarding the file operations from the guest to the host}. Our
solution to this challenge contributes two novel techniques: a virtual
memory technique, called \textit{hybrid address space}, that enables efficient
cross-domain memory operations, and the \textit{dual thread} technique
that efficiently leverages hypercalls to forward the operations and
improve concurrency in cross-domain operations.

Devirtualization currently supports four important classes of I/O devices for
personal computers using the same CVD implementation with fewer than 4900
LoC, of which about 300 are specific to each class: GPU, input
devices, such as mouse and keyboard, camera, and audio devices, such as
speaker.
We note that GPU has not been amenable to virtualization due to
its functional and implementation complexity. Yet, devirtualization
easily virtualizes GPU of various makes in laptop and desktop
computers with full functionality and adequate performance for multiple
guests.

We report a comprehensive evaluation of devirtualization. Our
evaluation shows that devirtualization requires low
development effort to support various I/O devices, easily shares
the device between the host and the guests, supports legacy devices, and is portable across different
versions of Linux. For interactive devices, such as input devices, camera, and speaker,
devirtualization achieves performance indistinguishable from that of
native by human user. For GPU, devirtualization achieves
close to or even higher than 60 frames per second (the display refresh
rate) on average for 3D HD games (1152$\times$864), even under stress
test by standard test engines.

We designed devirtualization for Unix-like OSes, such as Linux
distributions, Mac OS X, Android, and iOS, because they
constitute a large number of the installations on modern personal
computers, especially smartphones and tablets. However, we believe that
with proper engineering, devirtualization can also be useful for other
OSes, such as Windows, that also abstract several I/O devices with device files. 

Unfortunately, devirtualization is not universal and cannot virtualize
all devices, such as network and block devices. This is
because applications interface to these devices is sockets and file
systems, and not device files.
Fortunately, good solutions already exist for virtualizing these
devices~\cite{Russel2008}, as they have been critical for data centers
and basic use of VMs in personal computers.

\note{what other limitations?}

Devirtualization
introduces the device file interface between the guest and the host, which may be
abused by malicious guest applications. We are currently employing
techniques to guarantee isolation
between the system core, e.g., the host and the hypervisor, and the guests.
\S\ref{sec:discussion} elaborates more on this issue.



\section{Background}
\label{sec:background}

\begin{figure}
\centering
\includegraphics[width=0.9\columnwidth]{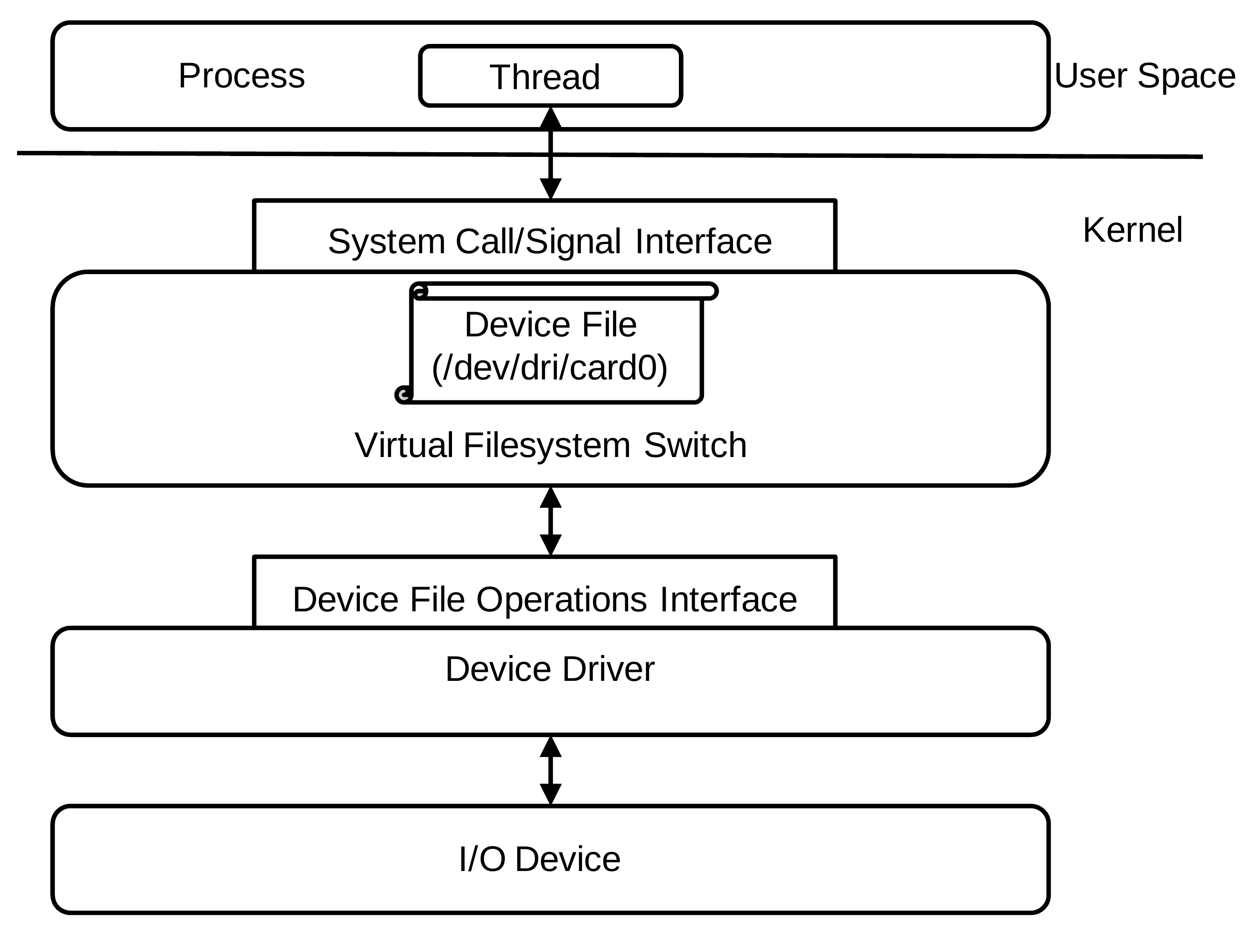}
\caption{The simplified I/O stack in Linux
}
\label{fig:boundary}
\customvspace
\end{figure}

Devirtualization targets virtualizing I/O devices for whole system
virtualization. It currently supports hosted hypervisors, where the hypervisors
runs inside a host OS, such as VMware Workstation.
The principle and design of devirtualization, however, apply to bare-metal
hypervisors equally well.

\subsection{I/O Stack and Devices Files}
\label{sec:device_files}

Devirtualization virtualizes I/O devices by virtualizing device files.
Figure~\ref{fig:boundary} shows a simplified I/O stack in Linux.
A process thread issues a file operation by calling the right system
calls to operate on the device file; these system calls are handled by
the Virtual Filesystem Switch (VFS), which invokes the \textit{file
operations} implemented by the device driver, e.g., {\tt read} and {\tt
memory map}. The kernel exports device files to user space
through a special filesystem, e.g., {\tt devfs (/dev)} in Linux. 
Important file operations for I/O devices include {\tt read},
{\tt write}, {\tt poll}, {\tt notification}, {\tt memory map}, {\tt
page fault}, and {\tt I/O control}.


Threads are the execution units in the OS,
issuing the file operations. All threads of a process share the process
address space. Therefore, we use ``thread" when
discussing the execution of file operations, but use ``process" when
discussing memory operations.
 
%


To correctly access an I/O device, an application may need to know
the exact model or functional capabilities of the device. For example,
the X Server needs to know the exact model of the GPU in order to load the
correct libraries. As such, the device driver and the kernel collect
this information and export it to the user space, e.g., through special
file systems of {\tt procfs} and {\tt sysfs} in Linux.

\subsection{Memory Virtualization}
\label{sec:memory_virtualization}

The hypervisor virtualizes the physical memory
for the guest. This creates a challenging barrier for devirtualization when
file operations from the guest must be executed in the host. 

There are two popular memory virtualization solutions. First, recent
generations of micro-architecture provide hardware support for
memory virtualization, i.e., Two-Dimensional Paging (TDP) as exemplified
by Intel Extended Page Tables (EPT). For TDP, the hardware Memory
Management Unit (MMU) performs two levels of address translation from
guest virtual addresses to guest physical addresses and then to system
physical addresses. Second, without hardware support, the hypervisor can
leverage a technique called shadow page
tables~\cite{Kivity2007} that utilizes the only level of translation in
the MMU to directly translate from the guest virtual addresses to system
physical addresses. The hypervisor maintains the shadow page tables and
keeps them in sync with the guest page tables, incurring a non-negligible
performance overhead. 



\subsection{Hypercall}

A \textit{hypercall} causes a transition from a guest OS to the
hypervisor, similar to a system call that causes a transition from the user space to the
kernel. The guest can use hypercalls to request privileged services
from the hypervisor. In modern architectures with hardware support for
virtualization, hypercalls use an instruction, e.g.,
VMCALL in x86, to switch the execution mode from the non-privileged
mode of virtualization to the privileged
mode. 

While the hypercall is in flight, the guest remains blocked and
cannot execute. This creates a challenge for devirtualization, since CVD
needs to perform potentially lengthy file operations in hypercalls.

\section{Overview of Devirtualization Design}
\label{sec:design}

Devirtualization virtualizes I/O devices at the device file boundary. It
allows the guest threads to use the host device drivers with a thin
layer of indirection: a virtual device driver.

\begin{figure}
\centering
\includegraphics[width=1\columnwidth]{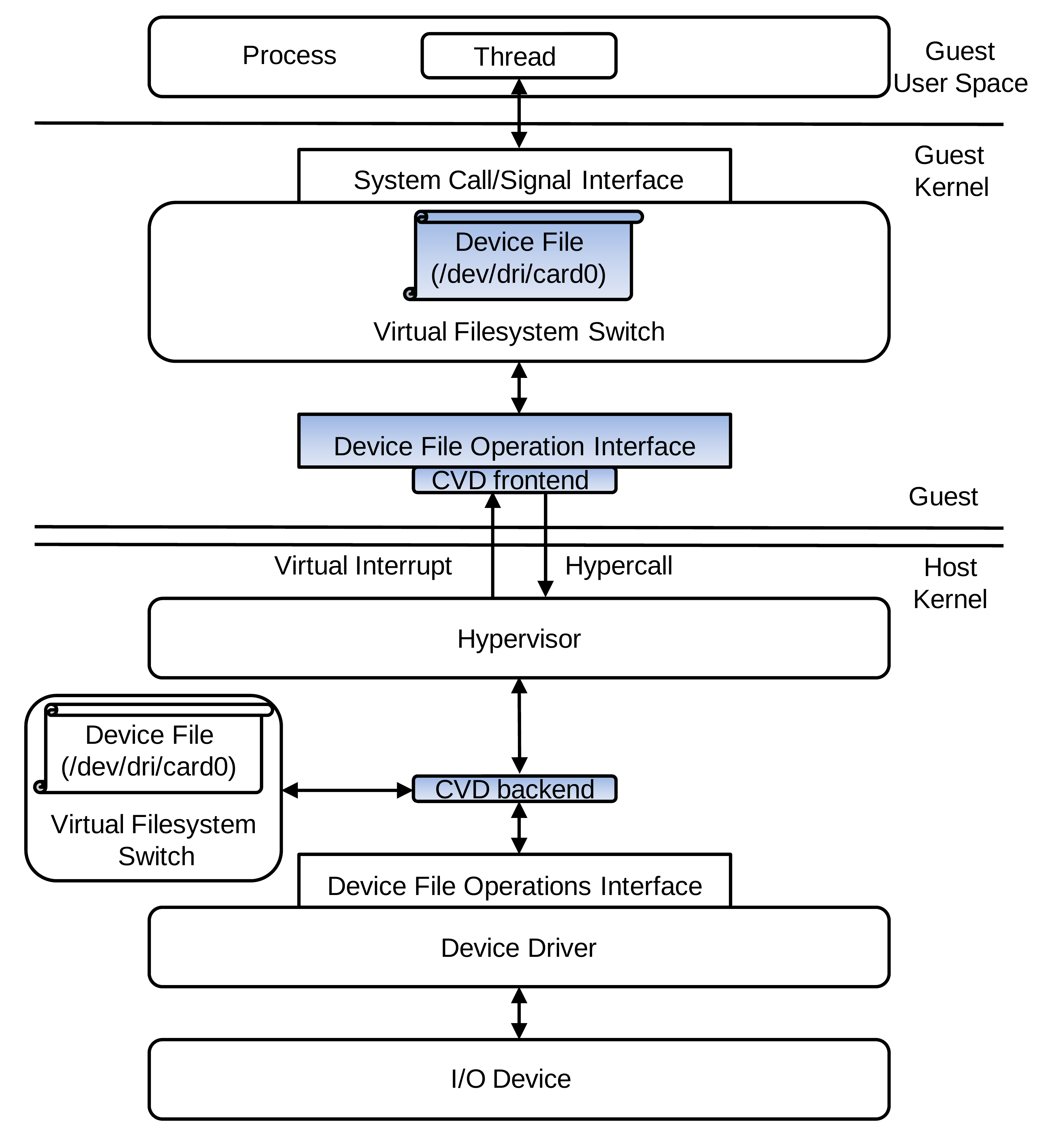}
\caption{Devirtualization architecture}
\label{fig:architecture}
\customvspace
\end{figure}

\subsection{Architecture}

Figure~\ref{fig:architecture} shows the architecture of devirtualization
with a single guest and a single I/O device. There are two components:
a {\em virtual device file} in the guest and the {\em Common Virtual
Driver} ({\em CVD}) with its frontend in the guest and the backend in
the host. To the guest,
the CVD frontend appears to be the device driver. However, instead of
servicing file operations, the CVD frontend forwards them to the CVD backend in
the host via hypercalls. The CVD backend then forwards these
operations to the host device driver. The results of the
file operation are returned to the CVD frontend and eventually to the
guest thread. 

When there are multiple virtual I/O devices, each one has its own
virtual device file but they share the CVD (hence the name
Common Virtual Driver). Note that this does not create
a single point of contention or failure for guest threads since the
CVD frontend and backend do not have active components, and their
routines are re-entrant and executed independently in the context of
the guest threads and their dual threads in the host
(\S\ref{sec:dual_thread}). 

\candidate{
\note{candidate for removal}
When a guest thread opens
a device file, the guest VFS creates a file handle data structure in the
kernel and returns a file descriptor to the guest thread (not shown in
Figure~\ref{fig:architecture}). Similarly, the CVD backend opens and
maintains a file handle in the host that mirrors the one in the guest,
and returns a file descriptor to the CVD frontend. 
} 

Devirtualization uses {\em virtual interrupts} to communicate from the
CVD backend to the frontend. For example, when a guest thread requests
{\tt notification} from a devirtualized device, the CVD backend informs the
CVD frontend of new events with an interrupt, and the CVD frontend then
signals the guest thread. Other uses of interrupts are for the
dual thread technique (\S\ref{sec:dual_thread}) and GPU sharing policy
implementation (\S\ref{sec:graphics_multiplexing}). For the CVD frontend to
be able to infer the purpose of each interrupt, the CVD backend either uses
different interrupt lines or writes integer arguments to a shared memory
page that can be read in the interrupt handler by the CVD frontend.

Devirtualization also extracts device information and exports it to the
guest OS by providing a small kernel module for the guest to load.
These {\em device info module} are small and easy to
develop, e.g., 100 and 50 LoC for GPU and camera, respectively.
\S\ref{sec:device_info_modules} provides more details on device info
modules.

\subsection{Portability}

We target devirtualization at running the same or different versions of
the same OS in the host and the guest.
We investigated the file operations interface of many versions
of Linux and observed the following: (i) the file operations that are
mainly used by device drivers, e.g., {\tt read}, {\tt memory map}, and
{\tt I/O control}, have been a part of Linux since the early days;
(ii) the complete set of file
operations have seen few changes in the past couple of years, i.e.,
three changes from Linux 2.6.35 (2010) to 3.2.0 (2012). These observations suggest that
supporting different versions of Linux in the host and the guest is
easy.
To demonstrate this, by adding only 14 LoC to the CVD, we
have successfully deployed devirtualization across two major versions of
Linux: version 2.6.35 for the host and version 3.2.0
for the guest, and vice versa.

If the guest has a different OS from that of the host, e.g.,
running a Windows guest on a Linux host, a translator is needed to
translate the file operations. This is part of our future work.

\subsection{Challenges}
\label{sec:challenges}

Devirtualization faces two important challenges
that stem from barriers enforced by virtualization hardware on
modern architectures.

First, some file operations, including {\tt read} and {\tt memory
map}, need the
host device driver to interact with guest process memory. However, the
guest virtual address space used by these operations is not valid
in the host. \S\ref{sec:hybrid_address_space} explains a novel
virtual memory technique, called \textit{hybrid address space}, to solve
this problem.

Second, hypercalls employed by devirtualization to forward file
operations block the guest while executing. This creates
significant problems for concurrency in devirtualization, degrading the
performance of other threads that do or do not use devirtualized devices.
\S\ref{sec:dual_thread} explains how the
dual thread technique mitigates this problem.

\section{Hybrid Address Space}
\label{sec:hybrid_address_space}


The hybrid address space allows the unmodified device driver in the host to
directly access the guest process memory as if it were accessing a host
process memory. This enables the host device driver to perform guest file operations,
such as {\tt read} and {\tt memory map}. ~\note{LZ: please
eliminate or minimize the use of devirtualization in this section and
next. First of all, devirtualization is a vague/abstract term and does
not point to any specific location in the system. Second, contributions
in this section and next are not specific to devirtualization. They are
the technologies that enable devirtualization. They can enable other
things as well. -- AAS: done }

\begin{figure}
\centering
\includegraphics[width=1\columnwidth]{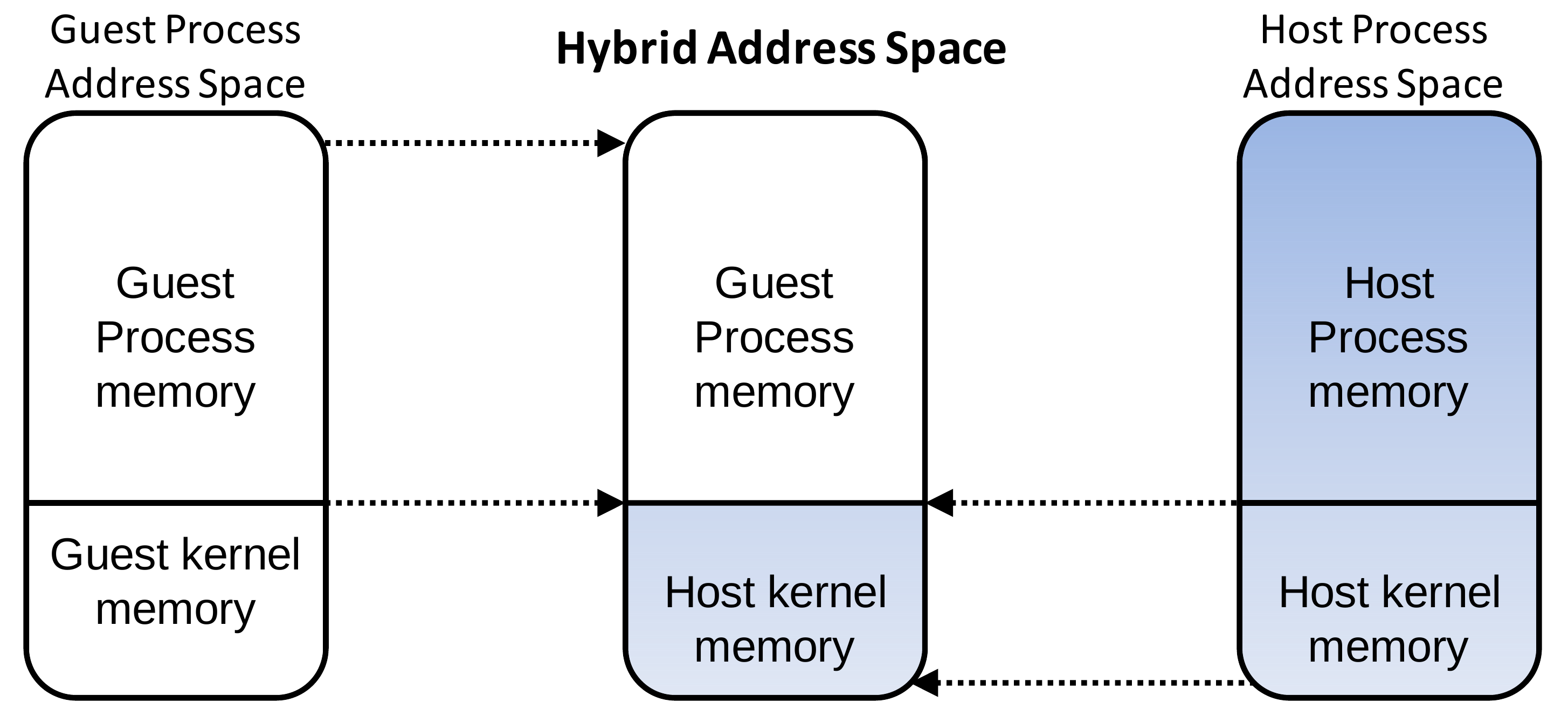}
\caption{Hybrid Address Space is a union of the guest process memory and
the host kernel memory.\note{align them in one horizontal line}}
\label{fig:hybrid_address_space}
\customvspace
\end{figure}

\subsection{Basic Idea}
\label{sec:has_design}

In modern operating systems, the address space of a process is the union
of the process memory and the kernel memory. As a result, when a process
thread makes a system call, the kernel, which executes in the context of
this thread, can access the process memory efficiently. For
brevity, we refer to this process and its thread as the current process
and thread, respectively. The hybrid address space is a similar union
of the guest process memory and the host kernel memory as illustrated by
Figure~\ref{fig:hybrid_address_space}. When the device driver needs to service a file operation forwarded from the guest, the
CVD backend\note{LZ: using ``devirtualization'' as the subject is very
vague. What exact in your design does the replacement? CVD backend? --
AAS: fixed} makes it `see' the hybrid address space rather than the host
address space,
\note{LZ: what is the host process? Did you mean the process of
the guest VM? -- AAS: fixed}
allowing the host device driver to
directly access the guest process memory. 
~\note{LZ: same problem as above -- AAS: fixed}

We provide both hardware and software realizations for hybrid address
space. In \S\ref{sec:has_comparison}, we
compare their pros and cons.


{\bf Example:} We use the following example to illustrate the role of the hybrid address
space\note{LZ: why do we need TWO examples? I felt one is enough. I also
felt the figure for {\tt read} is not worthwhile -- AAS: agreed.}:
A guest thread issues a {\tt memory map}
operation on the virtual device file to map the device or system memory in
the address space of its process. Through CVD, the operation is handed to the host
device driver, which creates the memory maps in the process portion of the
current host process address space. Since the hybrid address space is in
effect, the host device driver creates the memory maps for the guest
process. 



\subsection{Software Hybrid Address Space}
\label{sec:software_has}

Device drivers call certain kernel routines to interact with the process
memory. The CVD backend implements the hybrid address space in software by
redirecting and reimplementing these kernel routines to interact with
the guest (instead of the host) process memory. The software hybrid address does
not actually implement the address space in hardware,
but only creates an illusion of such an address space for the host
thread that is calling the device driver.

There are two categories of these kernel routines: the first category
enables the driver to read from and write to a user space buffer at
a given virtual address. In devirtualization, this virtual address is
a guest process virtual address, and is first translated to
a host physical or virtual address by the CVD backend, which can then write
to or read from the host address. If the size of the buffer is larger
than a page, the address translation needs to be performed once per
page,
since contiguous pages in the guest virtual address space are not
necessarily contiguous in host physical or virtual address spaces.

\note{AAS: I saw that this discussion was removed from the background
section. I add it briefly here since it is important for the reader to
be able to understand the sources of overhead in software hybrid address
space} 
The CVD backend translates a guest virtual address by first walking
the guest page tables in software to get the guest physical address. In
KVM, the guest physical address can then be simply
translated to its equivalent host virtual address in the guest VM process~\cite{Kivity2007}. With other
hypervisors, a software page walk of EPT or shadow page table will be
needed to finalize the translation.

The CVD backend\note{LZ: this is yet another example that you abuse
devirtualization to serve as an subject. The CVD backend does it. Now
the question is: is caching done per VM? -- AAS: fixed} also caches the translations
for future use, similar to how the TLB caches page table translations
in hardware. The caching is done per guest process. 
We use a simple
FIFO buffer with 10 entries for this cache, and measure its hit rate to
be about 90\%, even when running 3D HD games on a devirtualized GPU.
 
The second category of kernel routines enables the driver to map
a device or system memory page into the process address space at a given
virtual address. For these routines, the CVD backend creates the mapping
in the guest page tables and also in the shadow page tables or EPT,
depending on the memory virtualization type.\note{revise this sentence
-- AAS: done}

While fixing the shadow page tables or EPT is straightforward as they are
maintained by the hypervisor, fixing the guest page tables in the host
needs special attention. The CVD backend fixes the guest page table to map the guest virtual page to
a guest physical page. The guest physical page can be any arbitrary
page,
as long as it is not used by the guest OS. Using the hypervisor, the CVD
backend allocates several guest physical pages for this purpose.
Also, when fixing the guest page tables, the CVD backend might need to
allocate new guest physical pages to hold the new page table entries.
These pages must be recognized by the guest OS. Therefore, the CVD
frontend, upon initialization, allocates some pages for this purpose in
the guest and sends the address to the CVD backend for future use.

%
%
%

\note{LZ: this is a wrong way to end a design presentation. You shall
offer your insight into the pros and cons of the design or under what
circumstances this design will be particularly attractive. In this case,
it is particularly important because you are presenting two choices
here. You may choose to have this discussion at the end of Section
4 after you presented both software and hardware realizations. But here
you at least have to provide a pointer to that discussion so that
readers will postpone their curiosity. -- AAS: I added a sentence before
this section to tell the reader that we will have the comparison in the
end.}

\subsection{Hardware Hybrid Address Space}
\label{sec:hardware_has}

Alternatively, we can leverage the hardware MMU to realize the hybrid
address space\note{LZ: first notice how and why I removed
devirtualization in this sentence. Second after reading this sentence,
readers naturally wonder why one has to care about this realization
since you already have a software realization. Addressing my note at the
end of previous subsection may help but you need to read them through to
make sure it is smooth -- AAS: explained above}. In this realization, the CVD backend creates
a new page table for the guest process in the host, fixes that page
table to map the host kernel memory and the guest process memory, and
have the hardware MMU use this page table when the device driver is
servicing a guest file operation. The hardware hybrid address space
cannot be used when the guest uses TDP because TDP uses two address
translation, but the host MMU
can only perform one.


To create such a page table, the CVD backend leverages the shadow page
table maintained by the hypervisor
to find the guest process memory entries, and uses the current host process
page table maintained by the host OS to find the host kernel memory entries. 



The hardware hybrid address space can be realized with little overhead for
three reasons. First, it suffices to create the first level of the page
table, i.e., \textit{top-level page table}, for the guest process since
the next levels already exist in the shadow page table and in the host
page table; and the top-level page table is not larger than a single memory
page.
Second,
for every guest process, the top-level page table only needs to be
created once since it does not change. Third, the overhead of switching
to the hybrid address space is only a fraction of that required for a complete
context switch.



In essence, the hardware hybrid address space allows the host device driver
to directly manipulate the shadow page table for the guest process.
Caution must be taken in the implementation, since shadow page tables and
normal OS page tables
have subtle differences, e.g., the use of trapping entries rather
than non-present entries in shadow page tables.
We handle these issues in our
implementation, but do not further discuss them due to space
constraints.

\note{LZ: Why? -- AAS: it's just a candidate for removal if I ran out of
space close to the deadline. Mainly because it is very detailed.}
Finally, when the host device driver updates the shadow page table for
a guest process, the guest page table for that process is not updated accordingly. This
may seem to corrupt the guest
process memory, as shadow page tables should be in sync with the guest
page tables, but it does not,
since all the guest process file operations are handled in the
host using the hybrid address space. 

\note{LZ: as I noted at the end of 4.2, you owe the readers a discussion
of pros and cons of the two realizations and under what circumstances
each of them works best. -- AAS: added}

\subsection{Trade-offs}
\label{sec:has_comparison}

Both realizations of hybrid address space have pros and cons. The
important advantage of the software hybrid address space is that it
support both TDP and shadow page tables
(\S\ref{sec:memory_virtualization}), whereas hardware hybrid address
space does not support TDP. It is known that TDP has noticeably higher
performance than shadow page tables and is therefore widely adopted. 

On the other hand, the main concern with the software hybrid address space
is performance, mainly because of the extra overhead of software page
walks and multiple address translations for large buffers
(\S\ref{sec:software_has}), whereas in hardware hybrid address space,
these operations are as fast as native.
Fortunately, as we show in \S\ref{sec:gpu_benchmarks}, software and
hardware address spaces achieve close performance for a GPU benchmark,
but we note that this might not be true for other workloads or other I/O
devices. 

\section{Dual Thread}
\label{sec:dual_thread}

The goal of the dual thread technique is to efficiently forward the file
operations from the guest to the host and enable concurrency in
devirtualization, despite the challenges imposed by hypercalls as described
in \S\ref{sec:challenges}.

CVD employs hypercalls to forward guest file operations. 
However, hypercalls block the guest
\note{LZ: remove completely since it is not necessary. Apply
this principle to your text -- AAS: removed}while they execute. This creates two
important problems for concurrency in virtualized I/O devices\note{LZ:
see how I removed devirtualization}. First, blocking the
guest degrades the
performance of other threads in the guest that are not
even using the devirtualized I/O devices. Second,
hypercalls serialize the file operations, degrading the performance of
a virtualized I/O device if multiple virtualized devices are used
concurrently.

\begin{figure}
\centering
\includegraphics[width=1\columnwidth]{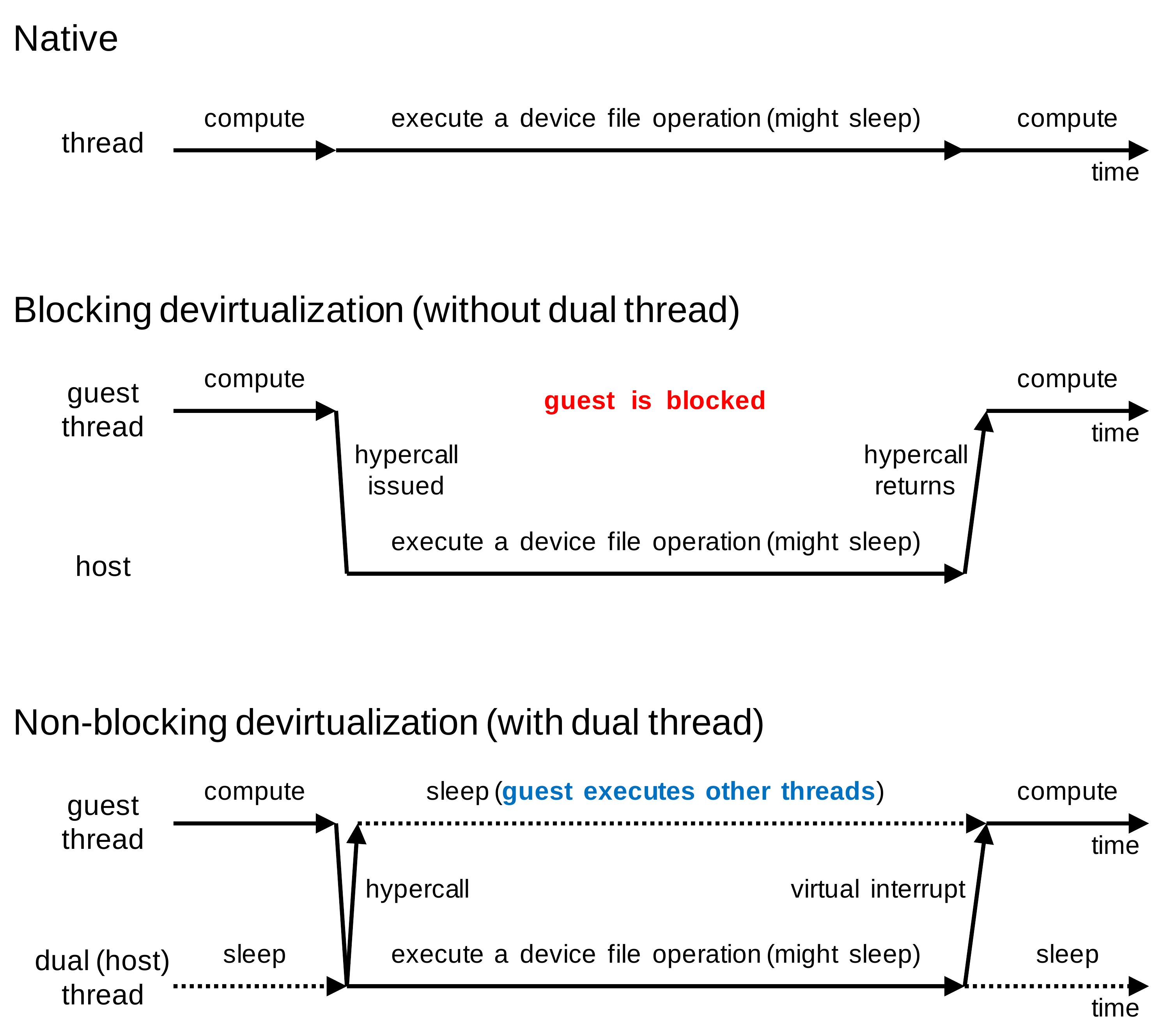}
\caption{The dual thread technique
}
\label{fig:dual_thread}
\customvspace
\end{figure}

We employ a technique called \textit{dual thread}\note{LZ: (1) dual
or mirror or what? I don't feel dual is a particular good choice of
word. (2) Why ``thread''? Insofar you have been using ``process'' in
this paper. Why suddenly switch to ``thread''? I understand that you use
thread to avoid the overhead of process. You need to bring this upfront.
For example, see my revision. I used ``another thread in the same
process''. Revise and polish. -- AAS: I fixed this inconsistency between
thread and process throughout the paper. Now I feel the text is much
more accurate technically. As I have explained in the background section
as well, thread is the unit of execution, but all threads of the same
process share the address space. Therefore, whenever we're talking about
execution, we use 'thread' and whenever we are talking about memory, we
use 'process'. This should clarify the ambiguity in this section as
well. Also, please read the next few paragraphs as well where I explain more
about the dual thread and its process, etc.} to solve these problems. As
illustrated in Figure~\ref{fig:dual_thread}, when the CVD backend
receives a file operation with a hypercall, instead of executing the
file operation in the context of the hypercall thread\note{AAS: I fixed
this revision here. The thread executing the hypercall is not sleeping
therefore CVD does not wake it up},
the backend wakes up another host thread to execute the operation, but
returns immediately from the hypercall so that the guest can resume
execution. We refer to this host thread as the dual thread of the guest
thread that issues the operation\note{LZ: apparently you named it
based on the relationships between the guest thread and the newly
created thread. You could also name it based on the relationships
between the host thread and the newly created thread. -- AAS: I'm fine
with other names if there's a good suggestion}. The dual thread
is spawned the first time that the guest thread issued a file operation.
While the dual thread is servicing the operation, the CVD frontend puts
the guest thread to sleep and allows the guest OS scheduler to schedule
other guest threads. When the operation is completed in the host, the
dual thread 
notifies the CVD
frontend using an interrupt. The CVD frontend then wakes up the guest
thread to continue.


The dual thread returns the result of the file operations on a memory
page shared with the guest thread. To this end, the CVD frontend allocates
one memory page for every guest thread that uses devirtualized I/O
devices and sends the (guest physical) address of this page to the
backend. The backend translates this address to a host address and
stores it for the dual thread to use.

The sleeping {\tt poll} operation is a good example to demonstrate the
benefits of the dual thread technique. This operation sleeps in the
kernel until an event is ready or there is a time-out. Without the dual
thread technique, the operation sleeps in the CVD backend in the
host, blocking the whole guest for potentially large intervals. However,
with dual thread, the hypercall immediately returns, and
the dual thread sleeps in the host instead. When an event
occurs, the host device driver wakes up the dual thread, which then
injects an interrupt into the guest to wake up the guest thread.

Finally, note that while the dual thread technique minimizes the
blocking of the guest, it can degrade the devirtualized I/O performance
for a single-threaded guest process (\S\ref{sec:gpu_benchmarks}). In such cases, this technique can be
disabled to improve performance. In the rest of the paper, we discuss
two operation modes for devirtualization: {\em blocking
devirtualization} that does not use the dual thread technique, and {\em
non-blocking devirtualization} that does. CVD can be easily configured to
operate in different modes for different processes or I/O
devices. \note{LZ: Why didn't you use the figure to explain how dual
thread technique works above? This paragraph is also placed at a weird
location. I would expect it comes at the end of the section. Right now,
it is interjected between the description of how dual thread works.
I will wait until you fix the comments in this section to continue. --
AAS: fixed}

\section{Implementation}
\label{sec:implementation}

We implement devirtualization for Linux/32-bit x86 platforms with the
KVM hypervisor. Our implementation works for both 2.6.35-24 and 3.2.0-52
Linux kernels running in Ubuntu 10.04 distribution. The implementation
is modular and can be revised to support other hypervisors and micro-architectures. Our implementation
virtualizes various GPUs, input devices, such as keyboard and mouse, camera, and speaker with less than 4900 lines of code;
only about 300 lines are specific to I/O classes. In particular, we have
tested our implementation with the following I/O
devices: discrete ATI Radeon HD 4650 GPU, integrated ATI Mobility Radeon
X1300 GPU on a Thinkpad T60 laptop, and integrated Intel Mobile GM965/GL960
\note{XXX}GPU on a Dell Latitude D630 laptop. The implementation
supports symmetric multiprocessor (SMP) guests.
 
\subsection{Common Virtual Driver (CVD)}
\label{sec:implementation_cvd}

The CVD is generic to all I/O devices, plays a critical role in
devirtualization, and constitutes a large portion of implementation. The
CVD frontend and backend consist of about 1030 and 2070 LoC respectively, and
are implemented as loadable kernel modules. The CVD backend has two
parts. The first implements the hybrid address space and the dual thread
technique and is therefore specific to the hypervisors and
micro-architecture. The second part interacts with the host OS only,
e.g., by calling the host device drivers. The CVD backend and frontend
also share a header file with about 560 LoC.

The CVD frontend uses hypercalls to forward a file operation to the backend.
Existing
Linux KVM/x86 hypercalls pass up to four arguments from the guest to the
host on virtual registers. We implemented a new hypercall to pass up to
6 arguments using the {\tt EAX}, {\tt EBX}, {\tt ECX}, {\tt EDX}, {\tt
ESI}, and {\tt EDI} x86 virtual registers. These six registers are
enough to pass the arguments of all file operations in one hypercall,
except for the {\tt page fault} operation, for which we use two
consecutive hypercalls. We added about 80 LoC to the guest and host
kernel for the new hypercall.



It is important to note that the Linux kernel employs generic class drivers to
unify all device drivers of the same class. For example,
the Direct Rendering Manager (DRM) driver is used to unify GPU drivers and the
event driver is used for input devices. These generic drivers create the
device files and export a file operation interface. They receive the
corresponding file operations and process and redirect them to the actual
device drivers through class-specific interfaces. In devirtualization,
the CVD
frontend plays the role of the generic driver and the actual device driver
altogether, and the CVD backend talks to the generic driver in the host.
Note that other non-devirtualized devices in the same guest can still
use their generic and actual drivers.

\subsection{Hybrid Address Space}
\label{sec:impl_has}



The hybrid address space enables cross-domain memory operations. For
the software hybrid address space, devirtualization redirects 9 Linux
routines to the CVD, e.g., {\tt insert\_pfn}, which maps a page to
a process address space. For redirection, devirtualization marks the
dual threads (or the thread executing the hypercall in blocking
devirtualization) and redirects the routines when called in the context
of marked threads. The marking is done by setting a flag in the
thread-specific structure {\tt task\_struct}. We have implemented the
software hybrid address space on the 3.2.0 kernel only; it requires
about 120 LoC in the host kernel and KVM.

To implement the hardware hybrid address space, we simply modify
the top-level page table as explained in \S\ref{sec:has_design}. We
use x86 with Physical Address Extension (PAE) paging in the host in our
current implementation. With PAE paging, the top-level page table is
called the Page Directory Pointer Table Entry (PDPTE) and contains
4 entries, each of which maps one fourth of the address space.
In x86, the {\tt CR3} register holds a pointer to the page table of the
current process. The CVD frontend finds the PDPTE of the current
process in the host using the host CR3 and finds the PDPTE of the shadow
page table of the current guest process using the virtual CR3 register.
We have implemented the hardware hybrid address space on the 2.6.35
kernel only. It requires 350 LoC in the host kernel and KVM.

\subsection{Dual Thread}

Dual thread improves the concurrency in devirtualization. We faced one
important challenge in the implementation of the dual thread technique.
That is, when the memory operations are executed in the context of the
dual thread, it may not be possible to find the location of the guest
process page table by reading the virtual {\tt CR3} register. This is because the
guest thread may be preempted in the guest. To solve this, the
CVD backend stores the location of the guest process as soon as one of
its threads makes a hypercall and uses that for
future address translations and memory maps. To support this, we
added about 60 LoC to KVM.

\subsection{Device Information Modules}
\label{sec:device_info_modules}

While the CVD is generic,
devirtualization requires a small amount of I/O class-specific code
to provide information about virtual I/O devices for guest
applications (\S\ref{sec:device_files}). For this, devirtualization employs
small kernel modules, or \textit{device info modules}, for the guest to
load. 

Developing the device info modules is easy because the modules are
simple and not performance sensitive. The device info module
for GPUs has about 100 LoC, and mainly provides information, such as the
PCI slot number, manufacturer and device ID, and 256 bytes of PCI
configuration data. The device info modules for an input device, camera, and
an audio device require 60, 50, and 50 LoC respectively.

In addition, we also developed a module to create or reuse a virtual PCI
bus in the guest for devirtualized devices. This module has about 290
LoC and can be reused for a large variety of PCI devices, such as GPUs.
Our current implementation of the PCI
module requires a small modification (50 LoC) to the guest PCI
subsystem.

\subsection{Sharing Policy}
\label{sec:graphics_multiplexing}

\note{LZ: People will ask: how about other I/O devices? how much
work do you have to do to get them virtualized with multiplexing? It is
so obvious here you only talk about the GPU. -- AAS: I tried to address
this}

By virtualizing I/O devices at the device file boundary,
devirtualization readily allows for the concurrent use of the device by the
both host and guests. However, we need to define the policy on
how each device is shared.

In the case of GPU, we adopt the
foreground-background model. That is, only the OS (host or guest) that
is in the foreground renders to the GPU, while the other OSes pause.
To achieve this, we assign each guest to one of the virtual
terminals of the host, and the user can easily navigate
between them using simple key combinations. When a guest goes to
background (or foreground), the CVD frontend receives an interrupt from the backend and then
signals the graphics application, e.g., X Server, to pause (or resume) rendering. 
If the guest does not pause, the CVD backend can forcefully reject all the operations
from that guest, although we have not yet implemented this. 
Implementing the graphics sharing policy required adding
15 LoC to the DRM driver (\S\ref{sec:implementation_cvd}) in order to
notify the CVD backend of change of the foreground OS.

Similarly, input devices should only send {\tt notifications} to the
foreground OS. To achieve this, the CVD backend only injects {\tt
notification} interrupts to the foreground guest. Similar simple
policies can be added for other devices as well.

\subsection{Driver-Initiated Memory Maps}

We faced a unique problem when applying devirtualization to the Intel
GPU Linux driver, i.e., {\tt i915}. We believe that our solution
(explained below) applies to other rare similar
situations as well, but we have not yet faced any.

Memory maps is almost always initiated in the user space with
a {\tt memory map} file operation. The kernel then determines a virtual
address range and calls the device driver to create the map.
However, the {\tt i915} driver initiates a memory map in the kernel
and as a result of an {\tt I/O control} operation (the rationale for
this design is explained in~\cite{Intel_GEM}). In devirtualization, the
host kernel cannot determine the guest virtual address range for the
memory map. Therefore,
the CVD backend records the map request, fails the file operation to go
back to the guest, allocates the virtual address range in the guest, and
re-executes the file operation, all hidden from the guest thread. Since
the first failed operation does not alter the state of the device or the
driver, it can be safely re-executed.

\section{Evaluation}
\label{sec:evaluation}

Using the implementation described above, we evaluate devirtualization
and show that it requires low development effort to support various I/O
devices, effectively shares the devices between the host and the guests,
supports legacy devices, and is portable across different versions of
Linux. For interactive devices such as input devices, camera, and
speaker, devirtualization achieves performance indistinguishable from
that of native by human user. For GPU, devirtualization achieves close
to or even higher than 60 frames per second (the display refresh rate) on
average for 3D HD games, providing a similar interactive user experience to
the native. 
\note{LZ: revise per my comment in the intro -- AAS: done}.

Unless otherwise stated, we use the following setup for all results. We
use a Dell
660s desktop using a quad-core Intel Core i5-3330s and 8GB of memory.
For I/O devices, we use a Radeon HD 4650 GPU, Dell mouse,
Logitech camera, and the Intel on-board sound card for speaker. We
configure the
guest with one virtual CPU and 1GB of memory. It uses TDP
(\S\ref{sec:memory_virtualization}) for its memory virtualization, and
therefore we configure devirtualization to use the software hybrid address
space. We compare the performance of a devirtualized I/O device with the
native device in all our measurement. We also report measurements for
the graphics virtualization solution from the VMware Workstation
9 hypervisor. 

\subsection{Non-Performance Properties}
\label{sec:eval_benefits}

Supporting new I/O devices with devirtualization is easy. For example,
we only needed to develop small device info modules with about 50 LoC
each to virtualize the camera and the speaker. It took only a few
person-hours to implement each of these modules.

Devirtualization easily supports the sharing of I/O devices between guests
and the host. For example, we are able to effectively share the GPU
between the host and two guests. One guest runs a 3D HD game, while the
host and the second guest run OpenGL applications. According to
the devirtualization policy (\S\ref{sec:graphics_multiplexing}), only the
foreground OS (guest or host) interacts with the GPU while others pause.
We can easily switch between the host and guests in less than
a second\note{LZ: can we quantify the latency? -- AAS: It is possible
but requires some work, not sure if I can do it in the time left.}. More
guests can be easily added as well. Note that devirtualization cannot
support sharing if the device driver or the device do not. For example,
the camera driver only supports one application at a time. 

Devirtualization supports legacy devices. Unlike sophisticated
self-virtualized devices~\cite{VMDq,VGX}, none of the devices that we
have successfully virtualized so far are specialized for virtualization.

Finally, we are able to run devirtualization with the host and guest
running two different versions of Linux: version 2.6.35 released in 2010
for the host and version 3.2.0 released in 2012 for the guest, and vice
versa. To support this, we added 14 LoC\note{LZ: awkward ``less than 15
LoC''. Why don't you just give the exact number of LoC? -- AAS: fixed}
to the CVD to update the list of file operations based on the new
kernel.

\subsection{Devirtualization Overhead Breakdown}
\label{sec:overhead_breakdown}

We look into the sources of overhead that degrade the devirtualization
performance compared to the native. The main source of overhead in blocking
devirtualization is hypercalls. Hypercalls incur two forms of
overhead. First, they add latency to each file operations. Second, they
increase the load on the CPU and pollute the cache~\cite{Gordon2012}.
While the latter is hard to measure, we measure the added latency using
a simple OpenGL application that draws a teapot and updates the screen
as fast as possible (\S\ref{sec:gpu_benchmarks}). We measure how long
it takes to issue all the file operations for drawing the teapot. Our
results show that blocking devirtualization executes these operations in
3.4 ms, while the native does so in 2.5 ms.

In addition to the overheads incurred by hypercalls\note{LZ: did you
mean non-locking devirtualization also suffers from the overhead of
hypercalls? -- Yes, it suffers from the two overheads of hypercalls in
the previous paragraph}, non-blocking devirtualization also suffers from the
overhead of the virtual interrupt (\S\ref{sec:dual_thread}), which 
takes an overage of about 44 $\mu$s. 
With this extra overhead, non-blocking devirtualization takes about 10.6
ms to execute the same file operations mentioned above.


\subsection{Performance of Interactive Devices}
\label{sec:eval_input}

We measure the latency of the devirtualized and the native mouse.
We note that the most accurate way of measuring the
latency of input devices is to measure the time it takes since the user
interacts with the device, e.g., the user moves the mouse, until the
time the effect of this event shows up on the screen. However, such
measurement is very difficult, especially in the case of mouse, which
generates many events in a short time. Instead, we measure the time from when the event is reported to the
host device driver by the mouse to when the {\tt read} operation
issued by the application reaches the host driver after the application
is notified of the event. Our results show that native and devirtualization
achieve about 16 $\mu$s and 73 $\mu$s of latency respectively, no matter how fast the
mouse moves. The extra overhead of devirtualization produces no
noticeable difference in our experiments. Much of this overhead is from
the virtual interrupt, which takes around 44 $\mu$s on average.
This suggests that the most effective way to suppress this overhead would be
a more efficient method for the host to signal the guest.

For the camera, we run the GUVCview~\cite{Guvcview} camera applications in the two
highest quality modes supported by our test webcam: 960$\times$720
resolution at 15 FPS, and 800$\times$600 resolution at
30 FPS. In both cases, there is no noticeable difference between the
native and devirtualized camera. For the speaker, we play the same
high-quality music file on both native and devirtualized speaker, and
achieve the same experience.

\subsection{Performance of GPU}
\label{sec:gpu_benchmarks}
\note{LZ: this subsection needs some structure. it is too long and flat
-- AAS: I put the hardware/software hybrid address space discussion in
a separate subsection. This indeed makes the evaluation section more
balanced since now there is one subsection for hybrid address space and
one for dual thread.}

We evaluate the performance of GPU devirtualization using interactive 3D
gaming and OpenGL applications. We use two 3D first-person shooter
games: \textit{OpenArena}~\cite{openarena} and
\textit{Tremulous}~\cite{Tremulous}, which are both widely used for GPU
performance evaluation~\cite{games}. For Tremulous, we use the Phoronix
Test Suite engine~\cite{phoronix}, a famous test engine that
automatically runs a demo of the game for a few minutes, while stressing
the GPU as much as possible. For OpenArena, we manually play the game,
therefore stressing the GPU less. We test the games at all supported
resolutions by the game or the test engine. For all our GPU evaluations, we report the standard
FPS metric. We also disable the GPU VSync feature, which
would otherwise cap the GPU FPS to 60 (display refresh rate) for a
smoother rendering. 

\begin{figure}[t]
\begin{minipage}[b]{0.49\columnwidth}
\centering
\includegraphics[width=1.1\columnwidth]{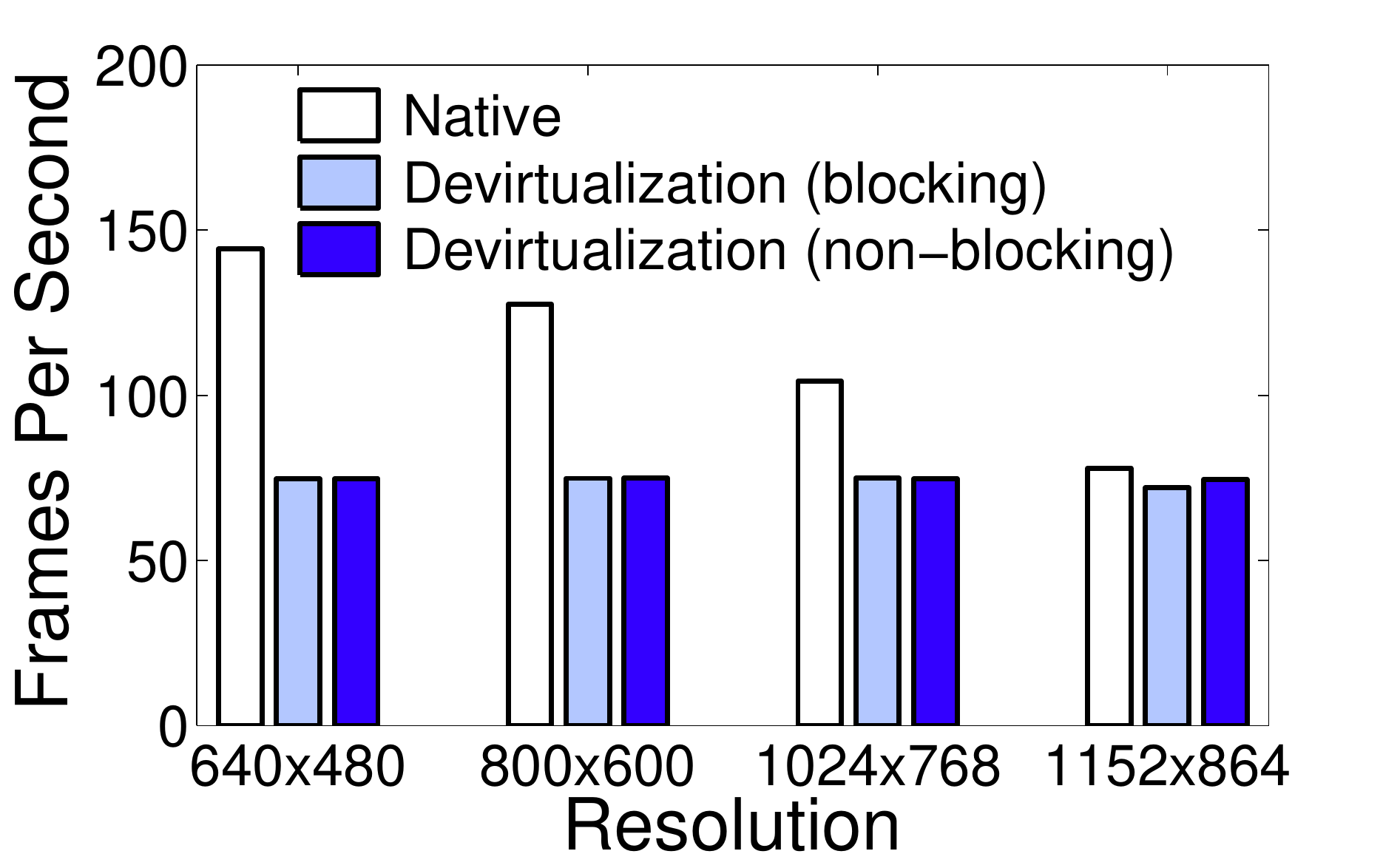}
\end{minipage}
\begin{minipage}[b]{0.49\columnwidth}
\centering
\includegraphics[width=1.1\columnwidth]{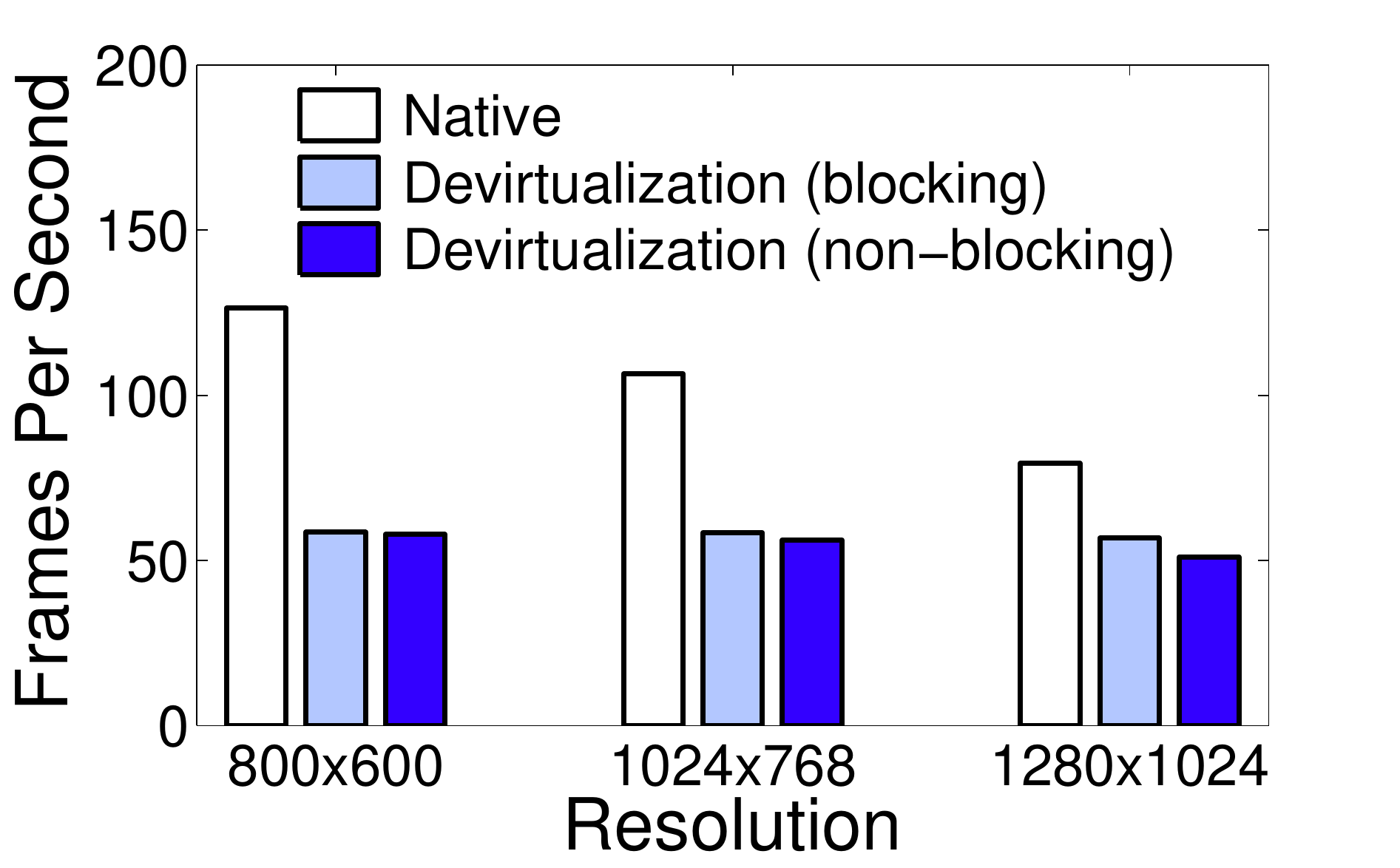}
\end{minipage}
\caption{FPS for running 3D interactive games of OpenArena (Left) and Tremulous (Right) at various resolutions supported by each game.}
\label{fig:games}
\end{figure}

Figure~\ref{fig:games} shows the results. There are two important
observations. First, in all scenarios, devirtualized GPU can achieve
close to or even higher than 60 FPS on average, providing a similar
interactive user experience as the native.
Second, devirtualization can achieve very close performance to the native at
high resolutions, but can show a noticeable gap with the native at lower
resolutions. This is because devirtualization adds a constant overhead
to file operations (\S\ref{sec:overhead_breakdown}) regardless of
resolution. This results in a lower percentage drop in performance
compared to the native at high resolutions, where the GPU needs more time
to render each frame. Finally, our results show that blocking and
non-blocking devirtualization achieve the same performance. We believe
this is the artifact of the game engine design, which, for example, may
report the same number for a range of FPS
values~\cite{openarena_graphics}.


We use an OpenGL benchmark~\cite{vbo_va} to show that non-blocking devirtualization
can indeed harm the performance noticeably in some scenarios. 
The benchmark draws a teapot that
has thousands of vertices, normals, and polygons using the Vertex Buffer
Objects API of OpenGL, and updates the screen as fast as possible.
We run the application for 3 minutes and measure the FPS. Our
measurement shows that native, blocking devirtualization, and
non-blocking devirtualization achieve an average of 97, 79, and 31 FPS,
respectively. Much of the performance degradation in non-blocking mode
is due to the overhead of virtual interrupts. Therefore, a more
efficient way to notify the guest from the host can help improve
non-blocking devirtualization. To confirm this, we implemented
a preliminary prototype that uses polling in the CVD frontend to find
out the completion of an operation instead of interrupts, and we managed to
boost the performance of non-blocking devirtualization to 54 FPS for
this OpenGL benchmark. However, we need to do more investigation to
better understand the implications of polling. 

We also measure the performance of the VMware Workstation
9 virtualization solution for graphics and find it to be significantly
lower than devirtualization. We use an Ubuntu host machine and an Ubuntu
guest with the same properties as the guest in our setup. The VMware
workstation uses VMware SVGA II driver for 3D graphics support. For the
Tremulous 3D game benchmark, VMware Workstation 9 achieves an average of
4, 2.7, and 1.8 FPS at the resolutions reported in
Figure~\ref{fig:games}. We note that other VMware products have
reportedly higher 3D performance~\cite{Dowty2009}, but to the best of
our knowledge, the VMware Workstation 9 with SVGA II driver is the only
one that could be configured on a Linux host to provide 3D graphics.



\subsection{Hybrid Address Space Realizations}

We compare the performance of software and hardware hybrid address space
and show that both achieve almost similar performances. With the hardware
hybrid address space, the OpenGL benchmark (\S\ref{sec:gpu_benchmarks})
achieves an average 74 FPS, which is close to the 79 FPS achieved by software
hybrid address space. The small difference may be due to the overhead of
shadow page tables used by the guest in hardware hybrid address space.
As mentioned in \S\ref{sec:impl_has}, our implementations of hardware
and software hybrid address space approaches are on different Linux
kernel versions; therefore it is questionable whether they can be
directly compared. However, since the performance of the native GPU on these two
kernels for the same OpenGL application is very close (96 FPS for
2.6.35 vs. 97 FPS for 3.2.0), we believe that the comparison is valid.

\subsection{Dual Thread and Concurrency}
\label{sec:eval_concurrency}

We measure the effectiveness of the dual thread technique to improve concurrency. We run the camera in
the guest in different devirtualization modes, i.e., blocking and
non-blocking, and show its impact on (\textit{i}) a concurrent compile
benchmark that compiles a simple C++ code segment in a loop for 100
times, and (\textit{ii}) an OpenGL application that uses the
devirtualized GPU concurrently. We configure the camera at
800$\times$600 resolution at a low 5 FPS, since a lower FPS
results in longer-lasting {\tt poll} file operations and is more
destructive to devirtualization. We repeat each experiment for
both UP (uniprocessor) and SMP guests with
two virtual CPUs. SMP allows the applications to run completely
concurrently and not compete for CPU time. 

\begin{figure}[t]
\begin{minipage}[b]{0.49\columnwidth}
\centering
\includegraphics[width=1.1\columnwidth]{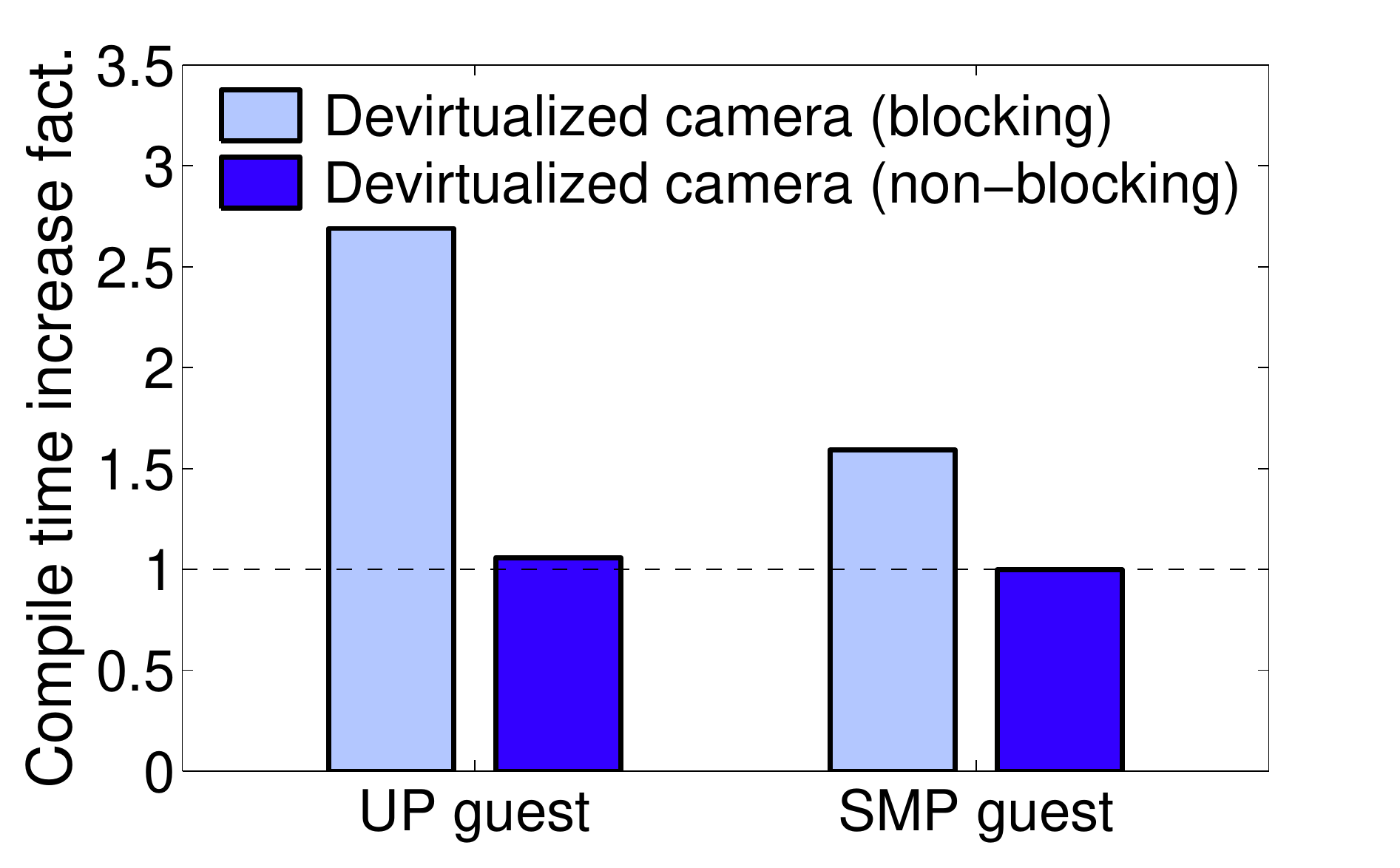}
\end{minipage}
\begin{minipage}[b]{0.49\columnwidth}
\centering
\includegraphics[width=1.1\columnwidth]{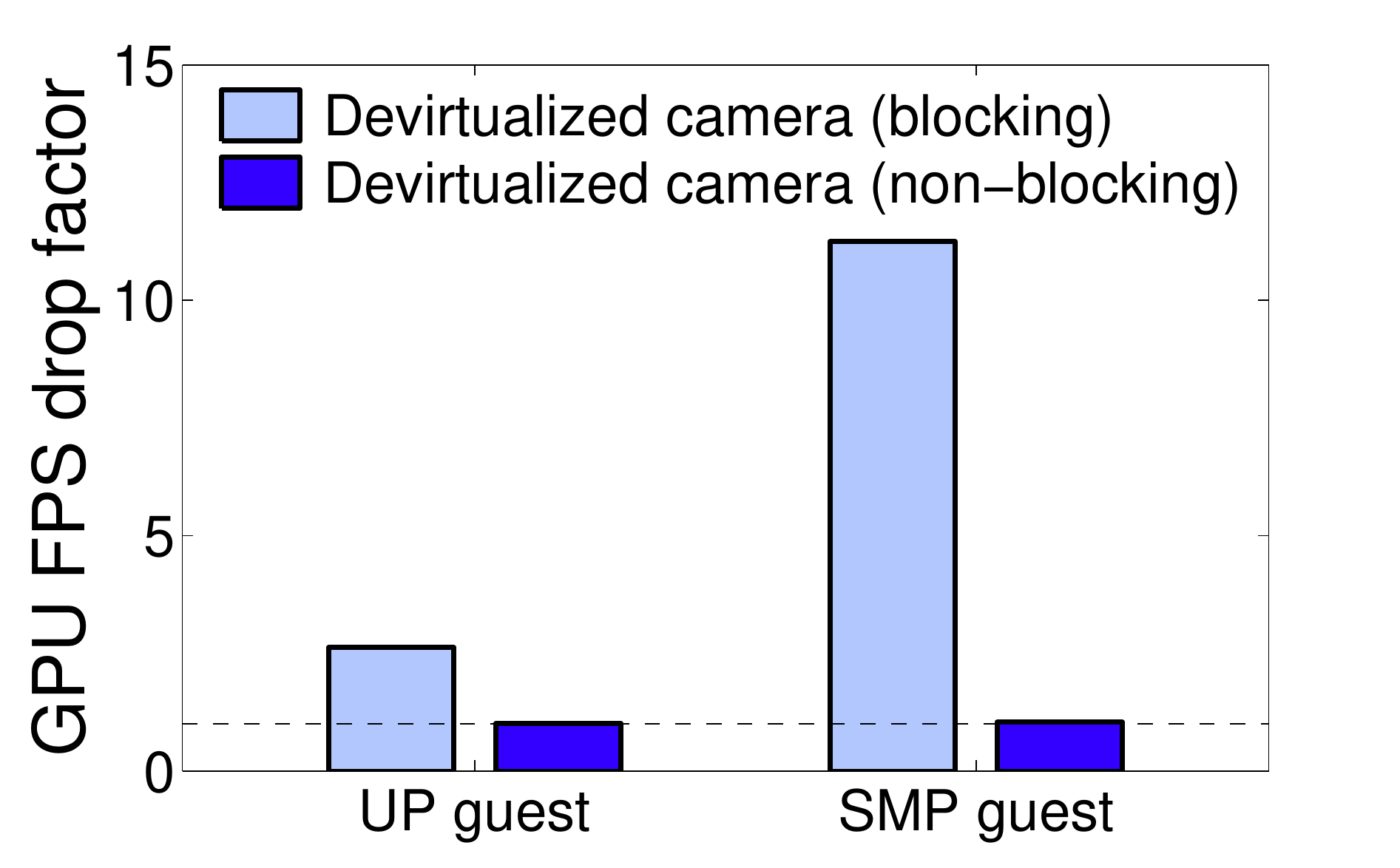}
\end{minipage}
\caption{Impact of a devirtualized camera at different devirtualization modes on a concurrent
compile benchmark (Left),
and on a concurrent devirtualized GPU (Right). The dashed line shows the no-impact value, or complete
concurrency.}
\label{fig:camera_concurrency}
\end{figure}

Figure~\ref{fig:camera_concurrency} (Left) shows the increase in compile
time as a result of the devirtualized camera. It shows that blocking
devirtualization increases the compile time by as much as 2.7$\times$, while
non-blocking devirtualization has almost no impact. The
slight increase of compile time for UP guest and non-blocking
devirtualization is the result of competition for CPU time.

Figure~\ref{fig:camera_concurrency} (Right) shows the drop in performance
(FPS) of the OpenGL benchmark as a result of the devirtualized camera.
It shows that blocking devirtualization can cause a 11.2$\times$ drop in GPU
performance, while non-blocking devirtualization achieves complete
concurrency with no performance drop. Surprisingly, blocking
devirtualization causes a much larger drop with the SMP guest than with
the UP guest. This is because with the SMP guest, the camera application
gets more CPU time and can continuously issue its file
operations, blocking the guest more often.

\section{Related Work}
\label{sec:related}

There are four major existing approaches toward I/O virtualization in whole
system virtualization. \textit{Emulation}~\cite{Sugerman2001} is
known to have poor performance. \textit{Direct I/O}~\cite{Willmann2008,
Ben-Yehuda2010, Gordon2012} provides close-to-native performance by
allowing the guest to directly own and access the physical devices;
however, it can only support a single guest OS. Moreover, direct I/O disallows the host
to use the device, which is a serious problem for GPU virtualization in
personal computers.
\textit{Self-virtualization} adds virtualization support to the I/O
device. Only sophisticated I/O devices, such as some network
interfaces, and a few high-end GPUs~\cite{VMDq,VGX,Dong2008}, use this approach, and therefore,
almost all legacy devices in personal computers are not supported.
\textit{Paravirtualization}~\cite{Dowty2009, Russel2008, Smowton2009,
Barham2003} employs paravirtual drivers in the guest and is most
related to devirtualization. Well-designed paravirtualization solutions
can achieve close-to-native performance. However, paravirtualization
requires significant development effort in order to support new
classes of I/O devices, and to support the full functionality of each I/O device.
For example, Xen3D only supports OpenGL
applications~\cite{Smowton2009}. In contrast,
devirtualization requires a one-time effort to support many classes of
I/O devices. It also allows threads \note{applications?} in the guest to use a device
driver in the host and enjoy the complete functionality of the
corresponding I/O device. Table~\ref{tab:io_virt} compares different I/O
virtualization solutions.



Cells~\cite{Andrus2011} employs user space virtualization (or operating
system-level virtualization) to run multiple virtual phones in one
Android smartphone. A virtual phone in Cells has its own user space, but
shares the kernel with other virtual phones. The use of user space
virtualization limits all virtual phones to have the same kernel and
provides rather weaker isolation between them compared to whole system
virtualization, as in devirtualization.

Some solutions provides graphics virtualization by remoting
OpenGL~\cite{Lagar2007, Parallels, Hansen2007, Virtualbox}, or
CUDA~\cite{Shi2009} APIs. Obviously, the applicability of these solutions are
limited to the specific graphics APIs.

Devirtualization allows the guest to reuse the device drivers in the
host.
It therefore provides a useful way to leverage legacy device drivers since driver
development is complicated and bug-prone~\cite{chou01sosp,swift03sosp}.
There have been related efforts in reusing legacy device drivers.
LeVasseur et al.~\cite{Levasseur2004} execute the device drivers in
a separate virtual machine and allow other guests to communicate
with
this VM for driver support. 
The driver domain model for
I/O paravirtualization in Xen~\cite{fraser2004safe} adopts a similar approach. 
However, the
guests must communicate with the driver
through an I/O class-specific interface, which requires nontrivial
development. In contrast, devirtualization builds the virtualization
boundary on device files, a common interface for I/O
devices, thus significantly reducing the development effort.

Dune~\cite{Belay12} provides direct access to virtualization hardware
features for user space processes. Dune processes use hypercalls to
issue system calls, which is similar to how devirtualization employ
hypercalls for file operations. However, Dune runs each process in
a separate VM and does not face the same blocking problem that
devirtualization does (\S\ref{sec:dual_thread}).

ELVIS~\cite{Gordon12} introduces exit-less notifications between the guest
and the host and is related to the dual thread technique.
However, unlike the dual thread, ELVIS requires a dedicated core in the
host for such notifications, which is acceptable for data centers, but
less so for personal computers. On the other hand, the dual thread
technique does not completely avoid exits like ELVIS, but it
reduces the duration of each exit caused by hypercalls.

\begin{table}[t!]
\begin{centering}
{\scriptsize
\begin{tabular}{|>{\centering}m{1.5cm}|>{\centering}m{1.2cm}|>{\centering}m{1.1cm}|>{\centering}m{1.6cm}|c|}
\hline
 & \bf{Perfor-} & \bf{Develop.} & \bf{Device} & \bf{Legacy} \\
 & \bf{mance}   & \bf{Effort}   & \bf{Sharing} & \bf{Support}\\ \hline
 \bf{Emulation} & \textcolor{red}{Low} & \textcolor{red}{High} & Yes & Yes \\ \hline
 \bf{Paravirt.} & High & \textcolor{red}{High} & Yes & Yes \\ \hline
 \bf{Direct I/O} & High & Low & \textcolor{red}{No} & Yes \\ \hline
 \bf{Self Virt.} & High & Low & Yes \textcolor{red}{(limited)} & \textcolor{red}{No} \\ \hline
 \bf{Devirt.} & High & Low & Yes & Yes \\ \hline
\end{tabular}
}
\caption{Comparison of I/O virtualization solutions\note{AAS: Since all
solutions can claim portability, I didn't add it to the table.}}
\label{tab:io_virt}
\end{centering}
\end{table}

\section{Discussion on Security}
\label{sec:discussion}

Devirtualization
introduces a new interface between the guest and the host, which may be
abused by malicious guest applications. Through the device file interface,
guest applications can either use the bugs in the device drivers (which
are known to be buggy~\cite{Ganapathi2006}) or the DMA capabilities of
I/O devices to write
to unauthorized locations in memory and break out of their VM isolation.
Moreover, guest applications might be able to prevent the device from
being fairly shared
with others. To combat these problems, We are working on
devirtualization to guarantee three
important forms of isolations:

{\bf Isolation of system core from malicious guest}: A guest must not be
able to tamper with the core components of the system, e.g., the
host and the hypervisor. To support this isolation, the device driver
and the device should be sandboxed. Existing work provides such
sandboxing techniques for both hosted and bare-metal hypervisors~\cite{Levasseur2004, Swift2003}.


{\bf Security isolation between guests}: For this, the CVD backend needs to
prevent one region of memory to be read or written by
two different guests.
 
{\bf Performance isolation between guests}: For this, the CVD backend
needs to be able to detect the abuse of the device by one guest, and then
boycott that guest by not forwarding the rest of its file operations.
However, a malicious guest might be able to push the device
to an unusable state before the boycott, in that case, the CVD backend can employ a device
recovery system, such as~\cite{Swift2004}.

\section{Conclusions}
\label{sec:conclusion}

We presented our attempt to provide an easy I/O
virtualization solution for whole system virtualization on personal
computers. Our solution,
called devirtualization, exploits a novel boundary that is narrow but common to
many I/O devices:
device files.  Using our design and implementation for Linux/x86
systems,
we are able to virtualize various GPUs, input devices, camera, and audio
devices with full functionality. Our measurements show that
devirtualization makes no user-perceptible difference, even when running
interactive 3D games in HD.  We consider this achievement remarkable,
particularly because GPU has been known to be difficult to virtualize.


\section*{Acknowledgements}

The work was supported in part by NSF Awards \#0923479, \#1054693, and \#1218041 and a gift from Nokia Research. The authors would like to thank
Kevin A. Boos and Jeffrey Bridge for their help with the implementation,
and Jon Howell from Microsoft Research for his useful comments.

\small
\bibliographystyle{plain}
\bibliography{iovirt}
\end{document}